# Test of the one-way speed of light and the first-order experiment of Special Relativity using phase-conjugate interferometers

### Ruyong Wang*, Yi Zheng and Aiping Yao
*St. Cloud State University, St. Cloud, MN 56301, USA*

With a Michelson interferometer using a phase-conjugate mirror (PCM) that reverses the uniform phase shift in a light path, we can conduct a first-order experiment of Special Relativity. Utilization of the PCM changes the basic concepts of an interference experiment. Placing a conventional partially reflecting mirror just in front of the PCM at the end of a light path, we can test the isotropy of the one-way speed of light in a system moving uniformly in a straight line and conduct the one-way Sagnac experiment. According to the reported phase-conjugate Sagnac experiment using a segment light path, we can expect that the phase shift is $\phi = 4\pi vL/c\lambda$ in the one-way Sagnac experiment with path length $L$ and speed $v$, even with an increasingly larger radius of the rotation. Based on these and the experimental fact of the generalized Sagnac effect, it is very important to examine whether there is the same phase shift for the test of the one-way speed of light and the first-order experiment using the PCM in a system in straight-line uniform motion. The sensitivities of these experiments are very high.

## 1. INTRODUCTION

The Michelson-Morley experiment [1] was perhaps the most important negative experiment in the history of science. Its null result led to Einstein's Special Relativity [2]. The experiment has been repeated many times in improved experimental conditions [3] and many kinds of Michelson-Morley type experiments have also been conducted [4,5]. All of them gave a null result. However, all these experiments were carried out in the laboratories which were stationary with respect to the earth. To fully verify the principle of relativity requires re-conducting the Michelson-Morley experiment at least in a reference frame moving relative to the earth [6,7], with which the principle of the constancy of the speed of light will also be examined. The Michelson-Morley experiment is a second-order experiment, i.e., the possible effect is proportional to $(v/c)^2$, where $v$ is the speed of the system. Therefore, conducting the Michelson-Morley experiment in a moving system would require a very high speed.

A first-order interference experiment has an intrinsic advantage over the second-order experiment. The possibility of the first-order experiment comes from the utilization of the nonlinear optical properties. While the conventional interferometers utilize beam splitters, mirrors and lenses that have linear optical properties, the utilization of the nonlinear optical properties brings a magnificent change to the interferometers. Phase-conjugate mirrors (PCMs) [8,9,10], the nonlinear optical devices, have an important property of phase reversal. Because of this property, the experiment utilizing the phase-conjugate interferometer can be the first-order. Besides, utilization of the PCM will totally change the basic concepts of the interference experiment. Unlike the traditional interference experiment with which the effective light paths constitute a closed loop, we can have a one-arm phase-conjugate interferometer. Actually placing a conventional partially reflecting mirror just in front of the PCM at the end of a light path, we can conduct the test of the one-way speed of light in a system in straight-line uniform motion. Because these experiments are the first-order, the sensitivities of these experiments are very high.

## 2. PHASE-CONJUGATE FIRST-ORDER EXPERIMENT

The Michelson-Morley experiment using the Michelson interferometer (Fig. 1) is a second-order experiment. For the time intervals that the light beam takes to travel from BS to $M_1$ and to travel back [11] we have

$t_1 = L/(c+v)$

and

$t_2 = L/(c\text{-}v)$

Hence,

$t_1 = L/[c(1+v/c)] = L/c - (L/c)(v/c) + (L/c)(v/c)^2 + \mathrm{O}[(v/c)^3]$

and

$t_2 = L/[c(1\text{-}v/c)] = L/c + (L/c)(v/c) + (L/c)(v/c)^2 + \mathrm{O}[(v/c)^3]$

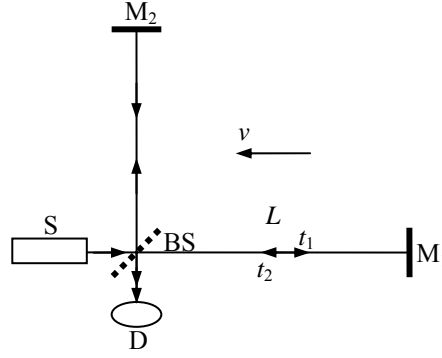

Fig. 1. The Michelson-Morley experiment. A light beam from the source S is divided into two beams, perpendicular to one another, by beam splitter BS. Two beams are reflected by mirrors, $M_1$ and $M_2$, and recombined by BS. The interference pattern is measured at detector D.

Neglecting terms of order higher than the second, the time interval that the light beam takes for a round trip is

$t = t_1 + t_2 = 2L/c + 2(L/c)(v/c)^2$

In term of the phase, from BS to $M_1$ the light beam experiences a phase shift of

$\phi_1 = 2\pi c t_1 / \lambda = 2\pi L/\lambda - (2\pi L/\lambda)(v/c) + (2\pi L/\lambda)(v/c)^2 + \mathrm{O}[(v/c)^3]$

Reflected from the mirror $M_1$, the light beam firstly experiences a phase change of one-half wavelength and then experiences a phase shift from $M_1$ to BS:

$\pi + \phi_2 = \pi + 2\pi c t_2/\lambda = \pi + 2\pi L/\lambda + (2\pi L/\lambda)(v/c) + (2\pi L/\lambda)(v/c)^2 + \mathrm{O}[(v/c)^3]$

Therefore, when the mirror is a conventional mirror, the total phase shift of the light beam is

$\phi = \phi_1 + \pi + \phi_2 = \pi + 4\pi L/\lambda + (4\pi L/\lambda)(v/c)^2$

neglecting terms of order higher than the second.

As a matter of fact, both $\phi_1$ and $\phi_2$ have the first-order term, $(2\pi L/\lambda)(v/c)$, but one is positive and another one is negative so they cancel each other. Therefore, the Michelson-Morley experiment with a conventional mirror is a second-order experiment. Besides, Lorentz proved that the interference experiments cannot detect the first-order effects [12] according to the classical theory.

The above derivations are based on the conventional interferometers and the phase shift of the light beam of a round-trip path is $\phi_1 + \phi_2$. However, if we have a method to change the positive phase shift to a negative phase shift, the first-order effect will augment each other instead of canceling each other. For example, if the mirror converts $\phi_1$ to $-\phi_1$, we will have a total phase shift of $-\phi_1 + \phi_2$. In this case the two first-order terms, $(2\pi L/\lambda)(v/c)$ in the Michelson-Morley experiment no longer cancel each other, but augment each other. Then we will have a first-order experiment. Actually changing a positive phase shift to a negative phase shift is what some PCMs can do.

The difference between a conventional mirror and such a PCM can be clearly shown in the experiments [13]. In the Michelson interferometer with the conventional mirror, a uniform phase shift in the path ($\Delta\phi$) will be doubled ($2\Delta\phi$) at the detector (Fig. 2a and 2b). However, if the conventional mirror is replaced by the PCM in the Michelson interferometer, the uniform phase shift in the path, $\Delta\phi$, will be cancelled at the detector. In fact, the light beam firstly experiences a phase shift $\Delta\phi$ when propagating east, the PCM reverses $\Delta\phi$ to $-\Delta\phi$, then the light beam experiences a phase shift $\Delta\phi$ again when it propagates back to the detector. The total phase shift becomes $-\Delta\phi + \Delta\phi = 0$. With the same reason, if a negative phase shift $-\Delta\phi$ is added in the incident path and a positive phase shift $+\Delta\phi$ is added in the reflected path (Fig. 3a), we will find a zero phase shift at the detector in the Michelson interferometer with a conventional mirror and a double phase shift at the detector in the Michelson interferometer with a PCM (Fig. 3b).

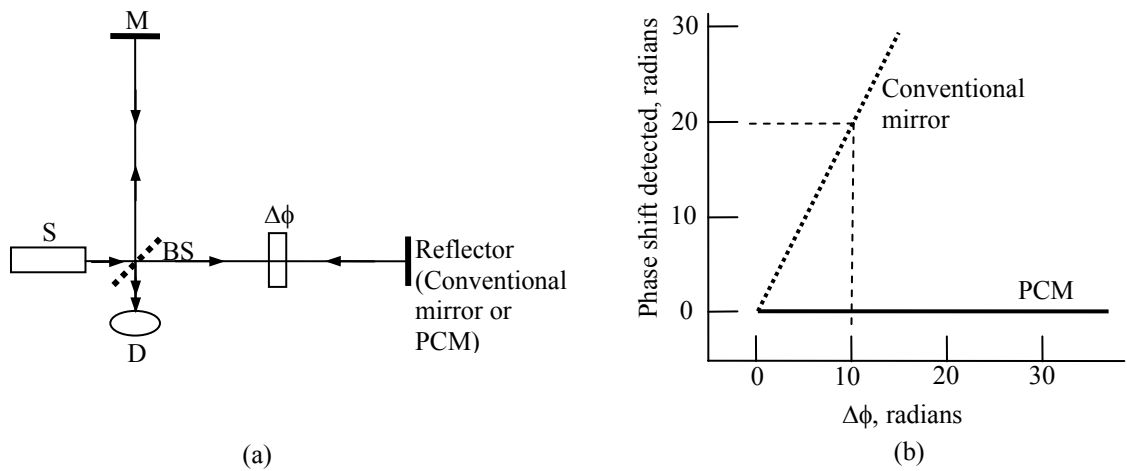

(a)

(b)

Fig. 2. The experiment shows the difference between a conventional mirror and a PCM. (a) The Michelson interferometer with a conventional mirror or a PCM. (b) Phase shift detected vs phase shift introduced.

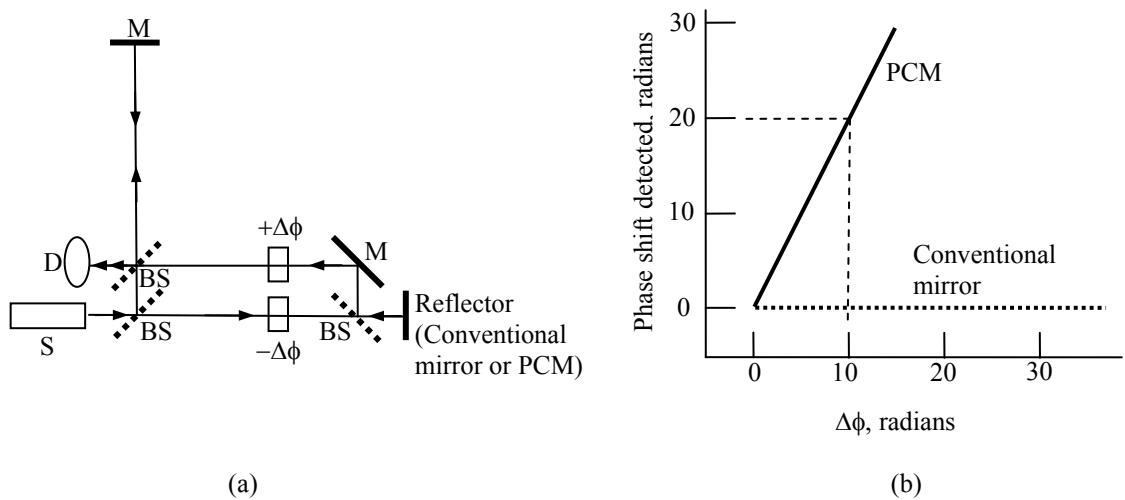

(a)

(b)

Fig. 3. Another experiment shows the difference between a conventional mirror and a PCM. (a) The Michelson interferometer with a conventional mirror or a PCM. (b) Phase shift detected vs phase shift introduced.

Therefore, if we conduct an experiment with a phase-conjugate Michelson interferometer shown in Fig. 4, the two first-order terms will augment each other:

$$\phi = -\phi_1 + \phi_2 = (4\pi L/\lambda)(v/c)$$

neglecting terms of order higher than the second. It is a first-order experiment. More generally, Lorentz's conclusion that the interference experiments cannot detect the first-order effects is not true for a phase-conjugate interferometer because the possibility of the phase reversal is not considered.

Generally speaking, changing $\phi_1$ to $-\phi_1$ is not the only way to have the first-order experiment. The experiment is valid as long as $\phi_1$ of the incident wave is not preserved as $\phi_1$ + constant of the reflected wave like in the case of a conventional mirror. For example, the phase shift can be changed to $-0.5\phi_1$, zero, or $2\phi_1$, etc. This is useful when we select proper PCMs for the experiments.

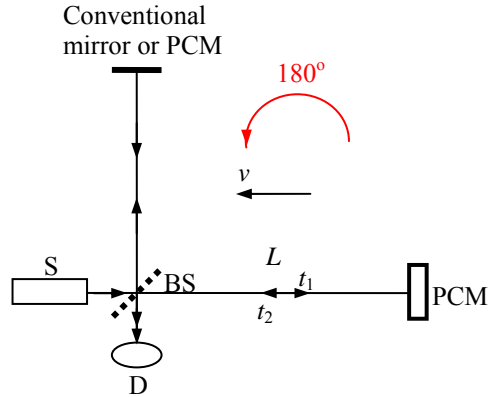

Fig. 4. The phase-conjugate first-order experiment.

Because only the first-order effect is concerned, the effect in the vertical arm of the Michelson interferometer, a second-order effect, is not considered. Lorentz contraction, a second-order effect, is not relevant in the first-order experiment. Moreover, the term related to the length of the horizontal arm, $2\pi L/\lambda$, does not appear anymore in the experiment shown in Fig. 4. Hence any distortion introduced to the horizontal arm such as vibration can be eliminated. Using the PCM in the vertical arm is not necessary, but is preferable to make the apparatus stable optically.

We can conduct the first-order experiment in several ways. The first one is to compare the phase shift of speed $v$ with that of zero speed, $\phi(v) - \phi(0) = 4\pi v L/c\lambda$. The second one is to compare the phase shifts before and after turning 180 degrees with speed $v$ (Fig. 4). Turning the phase-conjugate Michelson interferometer 180 degrees, the speed $v$ becomes $-v$ and the phase shift becomes $\phi - \phi_{(180)} = 8\pi v L/c\lambda$. Besides, we can conduct the experiment in a laboratory like the Michelson-Morley experiment did.

Assuming that the fiber technology is used and the detector has a sensitivity of $10^{-7}$ radians for the phase shift, with $L = 5$ m and $\lambda = 0.5$ μm, the experiment will be sensitive to a speed $v = 10^{-7} c\lambda/8\pi L = 0.12$ μm/s to check whether or not the phase shift exists.

## 3. PHASE-CONJUGATE SAGNAC EXPERIMENT

The Sagnac effect [14] shows that in a rotating closed path, two counterpropagating light beams take different time intervals to travel the closed path. For example, when the closed path rotates counterclockwise, the beam propagating counterclockwise takes a longer period of time than that of the beam propagating clockwise, and vise versa. The time difference is given by $\Delta t = 2R\Omega L/c^2$, where $R$ is the radius of the circular motion, $\Omega$ is the rotational rate and $L$ is the length of the path. The Sagnac effect is a fully verified experimental fact with numerous applications including highly precise fiber optic gyroscopes that are widely used in navigations. However, the arguments about its implications have never stopped and it is called an unsolved fundamental problem in physics [15]. A remarkable feature of the Sagnac effect is that all other effects related to the rotation, e.g., the centripetal force, will approach zero when the radius of the circular motion becomes increasingly larger while the speed of circular motion keeps constant; however, the Sagnac effect will not approach zero, but a finite value. Therefore, an apparent argument is that when the radius of the circular motion becomes increasingly larger, the circular motion approaches the linear motion and if the Sagnac effect still exists, i.e., the time difference still exists, does it contradict the principle of the constancy of the speed of light? The phase-conjugate Sagnac experiment on a segment light path [16], not the closed path like that in the most Sagnac experiments, makes this argument even more serious.

The experiment using a phase-conjugate Michelson interferometer is shown in Fig. 5. According to the Sagnac effect, from point A to point B, the light beam experiences a phase shift of

$$\phi_1 = kL - 2\pi R\Omega L/c\lambda$$

where $k = (2\pi n)/\lambda$ is the circular wave number.

In the return trip, the phase shift is

$$\phi_2 = kL + 2\pi R\Omega L/c\lambda$$

Because of the phase reversal on the PCM, the round-trip phase shift becomes

$$\phi = -\phi_1 + \phi_2 = 4\pi R\Omega L/c\lambda$$

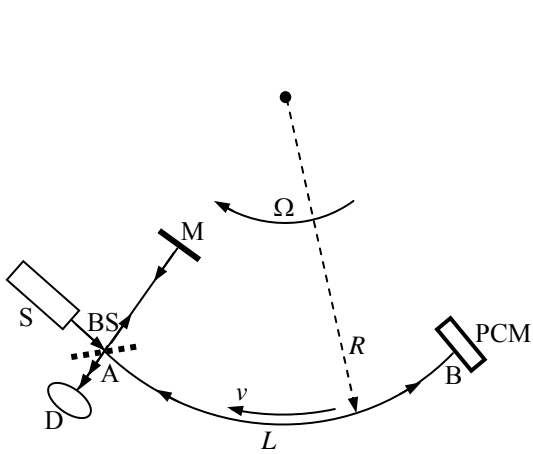

Fig. 5. The phase-conjugate Sagnac experiment.

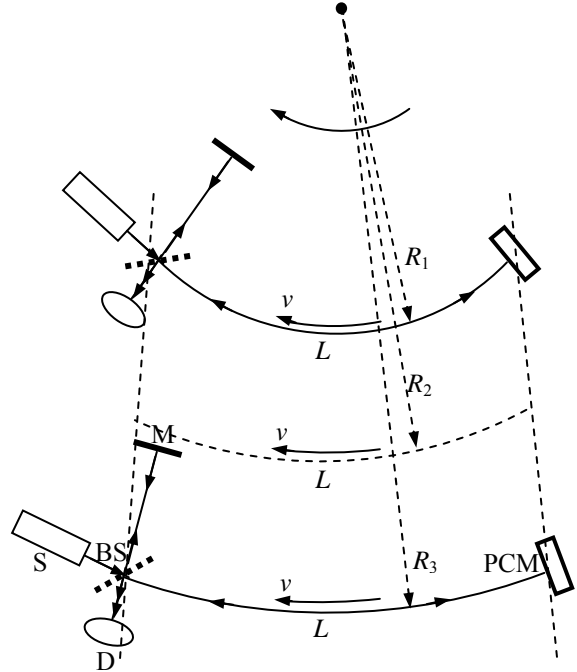

Fig. 6. The phase-conjugate Sagnac experiment with an increasingly larger radius of the rotation.

This experiment shows us two important points. First, it confirms the phase reversal of a PCM and demonstrates the Sagnac effect in an arc segment AB, not a closed path. Second, it gives us important implications as analyzed below. (Although in the experiment [16], the flexible fiber path was rotating and the other optical parts were not, in a similar experiment [17] all optical parts were rotating together.) The result, $\phi = 4\pi R\Omega L/c\lambda$, can be re-written as $\phi = 4\pi v L/c\lambda$ where $v$ is the speed of the moving arc segment AB. If we increase the radius of the circular motion as shown in Fig. 6, the arc segment AB will approach a linear segment AB, the circular motion will approach the linear motion, the phase-conjugate Sagnac experiment will approach the phase-conjugate first-order experiment as shown in Fig. 4, and the phase shift is always $\phi = 4\pi v L/c\lambda$. Therefore the result of this experiment indicates what we could expect from the phase-conjugate first-order experiment.

## 4. GENERALIZED SAGNAC EFFECT

That a segment in linear motion contributes a phase shift $\phi = 4\pi v L/c\lambda$ not only is a deduction from the phase-conjugate Sagnac experiment, but also is based on the experimental fact of the generalized Sagnac effect [18,19]: for any moving segment of a loop there is a phase difference, $\phi = 4\pi v L/c\lambda$, between counter-propagating light beams in the segment of length $L$ and speed $v$ whether the segment is in circular motion or linear motion. (Generally, the phase difference is proportional to a product of the moving velocity $v$ and the projection of the segment length $L$ on the moving direction, $\phi = 4\pi\boldsymbol{v}\cdot\boldsymbol{L}/c\lambda$.) The experimental fact was confirmed with different configurations, circles and conveyors, conveyors with different lengths, two-wheel and three-wheel conveyors, figure-8 and zero-area conveyors, and parallelograms (Fig. 7). Clearly, in the segment moving linearly, there is a phase difference $\phi = 4\pi vL/c\lambda$, between counterpropagating beams. Therefore, it is very important to see whether the phase-conjugate first-order experiment will give a non-null result, $\phi = 4\pi vL/c\lambda$.

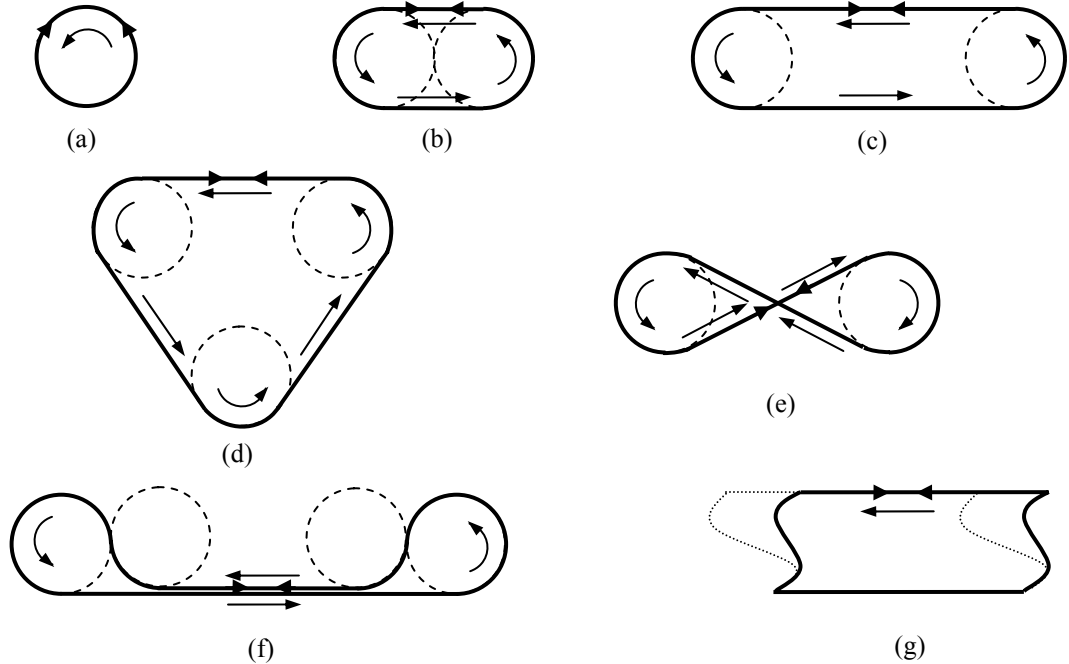

Fig. 7. Different configurations used in the experiments of the generalized Sagnac effect. (a) Circle. (b) Two-wheel conveyor. (c) Two-wheel conveyor with a longer path. (d) Three-wheel conveyor. (e) Figure-8 conveyor. (f) Zero-area conveyor. (g) Parallelogram.

## 5. TEST OF THE ONE-WAY SPEED OF LIGHT

The importance of the test of the one-way speed of light can never be overestimated because the isotropy of the one-way speed of light not only is the foundation of the principle of the constancy of the speed of light, but also is the core of the problem of simultaneity. It was stated by Einstein: "That light requires the same time to traverse the path $A \rightarrow M$ as for the path $B \rightarrow M$ is in reality neither a supposition nor a hypothesis about the physical nature of light, but a stipulation which I can make of my own freewill in order to arrive at a definition of simultaneity." ($M$ is the mid-point of line $AB$.) [20] The test of the one-way speed of light is difficult because the clock type of the experiments has the problem of the clock synchronization and the interference type of the experiments has the problem of the closed light path. Sometime it was even claimed that the one-way speed of light cannot be tested by the experiments [21].

For an interference experiment, the following concepts are always used explicitly or implicitly: the basic operating principle of an interference experiment is that a light ray from the source is divided into two rays 1 and 2 at point A (Fig. 8), the two rays traverse different paths I and II and they are brought to interference at another point B, the detector measures the phase difference between the two different paths [11]. It is also naturally thought that the effective light paths I and II for the interference experiment must constitute a closed loop and that the path from the source to the dividing point A does not have any role to the interference pattern measured at the detector. However, these concepts are true only for a traditional interference experiment and they are not necessarily true for a phase-conjugate interference experiment. Because a PCM has the property of phase reversal, we can have a one-arm phase-conjugate interferometer: placing a beam splitter $BS_1$, a conventional partially reflecting mirror, just in front of the PCM (Fig. 9a), or using a PCM with a partially reflecting coating (Fig. 9b), at the end of a one-way light path $L$. In this way, the light path $L$ actually plays two roles, one as a path with a conventional mirror and one as a path with a PCM. If there is a phase shift $\Delta\phi$ in this one-way light path $L$, the conventional mirror preserves the phase shift and the PCM reverses the phase shift. Therefore a detector near the end of the path will detect a double phase shift, i.e., $-2\Delta\phi$ (Fig. 9c), between the two beams that are from the conventional mirror and the PCM. Obviously, the concepts that the effective light paths must constitute a closed loop and the path from the source to the dividing point does not play any role to the interference pattern at the detector are not true anymore.

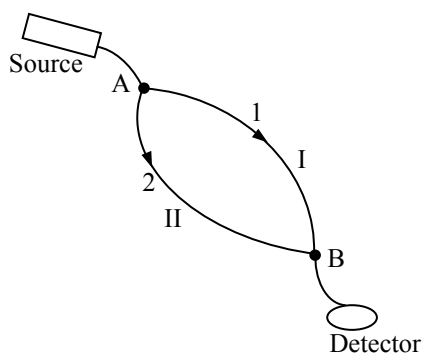

Fig. 8. The basic configuration of a traditional interference experiment.

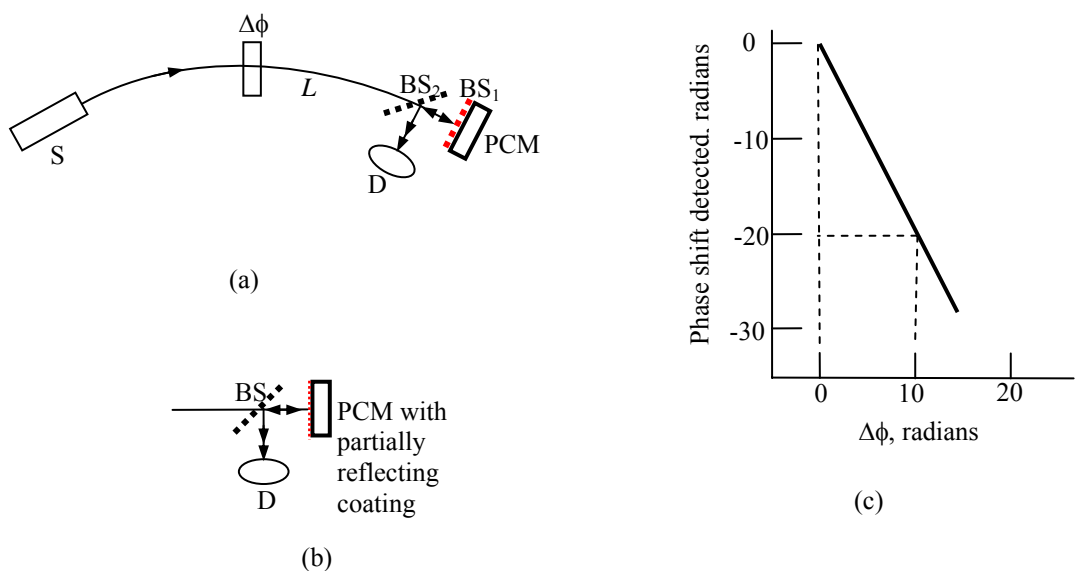

(a)

(b)

(c)

Fig. 9. One-arm phase-conjugate interferometer. (a) Placing a beam splitter, a conventional partially reflecting mirror, just in front of the PCM. (b) PCM with partially reflecting coating. (c) Phase shift detected vs phase shift introduced.

Therefore with phase-conjugate interferometers, we can conduct the tests of the one-way speed of light for a system moving uniformly in a straight line (Fig. 10a) and for one-way Sagnac effect (Fig. 10b). If there is a phase shift, $-2\pi vL/c\lambda$, in this one-way light path $L$ moving with speed $v$, the detector near the end of the path will detect a double phase shift, i.e., $4\pi vL/c\lambda$. A practical configuration for the experiment is shown in Fig. 10c.

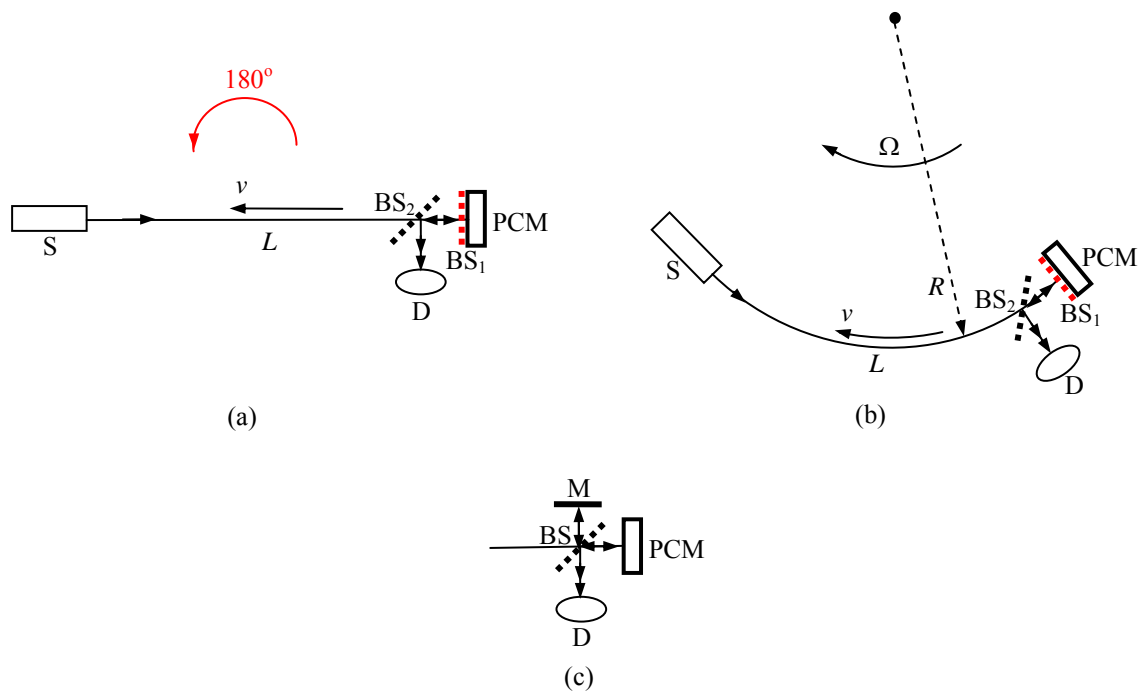

Fig. 10. Tests of the one-way speed of light with phase-conjugate interferometers. (a) Test of the one-way speed of light in a system moving uniformly in a straight line. (b) One-way Sagnac experiment. (c) A practical configuration.

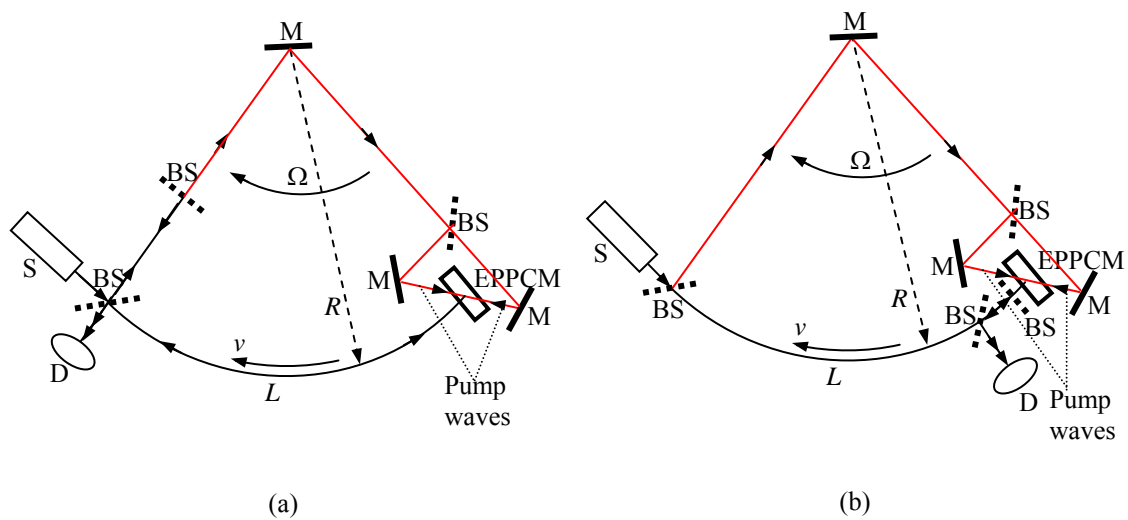

Fig. 11. The Sagnac experiments with an EPPCM. (a) Phase-conjugate Sagnac experiment. (b) One-way Sagnac experiment.

## 6. SELECTING PROPER PCMs FOR THE EXPERIMENTS

A proper PCM that reverses the uniform phase shift should be selected for the experiments. There are many types of the PCMs [22,23]. However, not all the PCMs have this phase reversal property. Those PCMs that are self-starting, for example the stimulated Brillouin scattering, have overall phase ambiguities that prevent compensation for optical path difference and they cannot provide the phase reversal [24].

The externally pumped PCM (EPPCM) with the four-wave mixing [25,26] can perform the phase reversal but the impact of the two pump waves should be studied. For the phase-conjugate Sagnac experiment or the one-way Sagnac experiment, pump waves pass in the radial directions (Fig. 11a and 11b) and there is no a phase shift due to the pump waves, so an EPPCM is suitable.

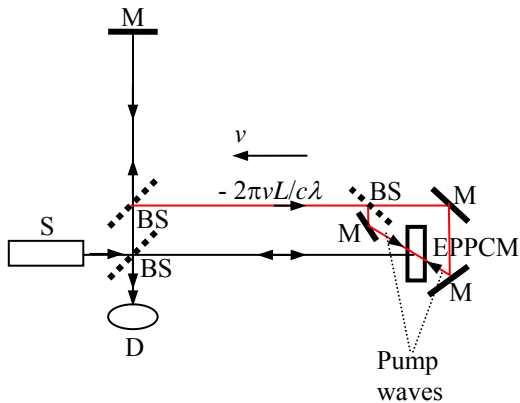

Fig. 12. The Michelson interferometer with an EPPCM and one laser.

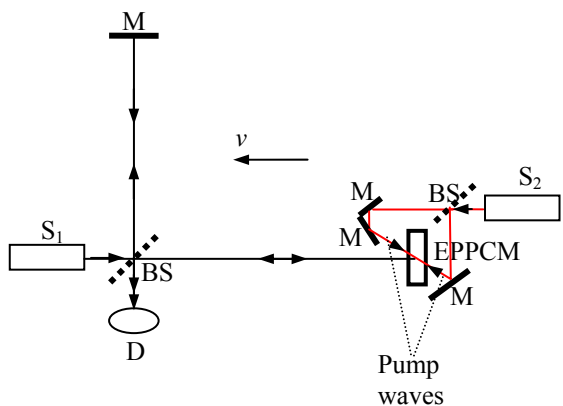

Fig. 13. The Michelson interferometer with an EPPCM and two separate lasers.

Let us consider the EPPCM for the phase-conjugate first-order experiment. As shown in Fig. 12, each of the two pump waves experiences a phase shift of $-2\pi vL/c\lambda$, which is the same as the incident wave. The PCM reverses the phase shift of the incident wave. However, because of the existence of the phase shift brought by each of the two pump waves [13], the phase shift of the reflected wave left the PCM is $2\pi vL/c\lambda - 2\pi vL/c\lambda - 2\pi vL/c\lambda = -2\pi vL/c\lambda$. Therefore, the reflected wave from the EPPCM in this configuration does not offer the desired phase shift and cannot be used for the first-order experiment. Similar to the configuration proposed for the test of the gravitational field [27] in which two EPPCMs have two separate lasers for pumps waves, a configuration in which the pump waves are from a separate laser is needed for the first-order experiment (Fig. 13).

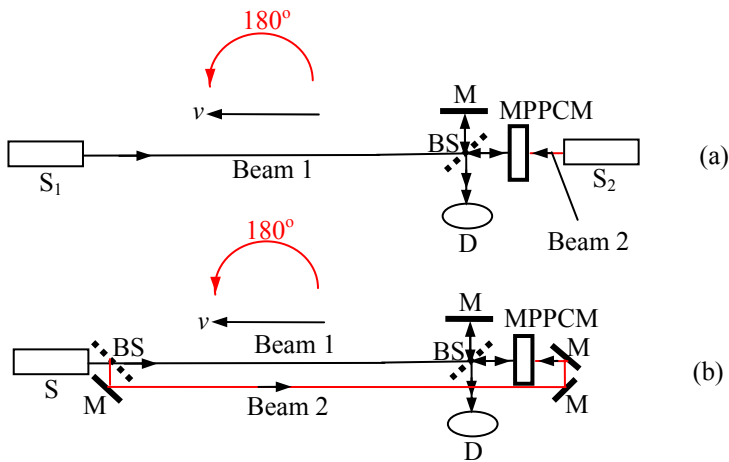

Fig. 14. Tests of the one-way speed of light with a MPPCM. Beam 1 is used and beam 2 is unused for the interference. (a) Two light sources. (b) One light source.

The mutually pumped PCM (MPPCM) [28,29] provides the phase reversal for the uniform phase shift [30]. Two configurations of using the MPPCM for the test of the one-way speed of light are shown in Fig. 14a and 14b, where one uses two light sources having the same frequency and another one uses only one light source.

## 7. CONCLUSIONS

The phase-conjugate Sagnac experiment can be repeated and the one-way Sagnac experiment can be conducted using the PCM. We can expect that the phase shift is $\phi = 4\pi v L/c\lambda$ in the one-way Sagnac experiment with path length $L$ and speed $v$, even with an increasingly larger radius of the rotation. Based on these and the experimental fact of the generalized Sagnac effect, it is very important to examine whether there is the same phase shift for the test of the one-way speed of light and the phase-conjugate first-order experiment in a system moving uniformly in a straight line. The sensitivities of these experiments are very high.

*E-mail address*: ruwang@stcloudstate.edu

**References**

[1] A. A. Michelson and E. W. Morley, *Am. J. Sci.* **34** (1887) 333.
[2] A. Einstein, *Ann. der Phys.* **17** (1905) 891.
[3] R. S. Shankland, *et al.*, *Rev. Mod. Phys.* **27** (1955) 167.
[4] H. Müller *et al.*, *Phys. Rev. Lett.* **91** (2003) 020401.
[5] P. L. Stanwix, *et al.*, *Phys. Rev. Lett.* **95** (2005) 040404.
[6] R. Wang, Z. Chen, X. Dong, *Phys. Lett.* **75A** (1980) 176.
[7] M. Psimopoulos, T. Theocharis, *Nature* **319** (1986) 269.
[8] B. Ya. Zel`dovich, *et al.*, *ZhETF Pis. Red.* **15** (1972) 160.
[9] R. W. Hellwarth, *J. Opt. Soc. Amer.* **67** (1977) 1.
[10] A. Yariv, D. M. Pepper, *Opt. Lett.* **1** (1977) 16.
[11] C. Møller, The Theory of Relativity, 2nd edition, Clarendon, Oxford, 1972.
[12] H. A. Lorentz, The Theory of Electrons and Its Application to the Phenomena of Light and Radiant Heat, 2nd edition, Dover, New York, 1952.
[13] R. W. Boyd, *et al.*, *Opt. Lett.* **12** (1987) 42.
[14] G. Sagnac, *C. R. Acad. Sci.* Paris **157** (1913) 708.
[15] J. P. Vigier, *Phys. Lett. A* **234** (1997) 75.
[16] P. Yeh, I. McMichael, M. Khoshnevisan, *Appl. Opt.* **25** (1986) 1029.
[17] I. McMichael, P. Yeh, *Opt. Lett.* **11** (1986) 686.
[18] R. Wang, Y. Zheng, A. Yao, D. Langley, *Phys. Lett. A* **312** (2003) 7. (Also see: http://arxiv.org/ftp/physics/papers/0609/0609222.pdf)
[19] R. Wang, Y. Zheng, A. Yao, *Phys. Rev. Lett.* **93** (2004) 143901. (Also see: http://arxiv.org/ftp/physics/papers/0609/0609235.pdf)
[20] A. Einstein, Relativity: The Special and General Theory, 15[th] Edition, Crown, New York, 1952.
[21] Y. Z. Zhang, Special Relativity and Its Experimental Foundations, World Scientific, Singapore, 1997.
[22] R. A. Fisher, Ed., Optical Phase Conjugation, Academic Press, New York, 1983.
[23] B. Ya. Zel`dovich, N. F. Pilepetsky, V. V. Shkunov, Principles of Phase Conjugation, Springer-Verlag, Berlin, 1985.
[24] R. A. Fisher, *Optical Phase Conjugation*, in Encyclopedia of Lasers and Optical Technology, edited by R. A. Meyers, Academic Press, San Diego, 1991.
[25] J. P. Huignard, J. P. Herriau, P. Aubourg, E. Spitz, *Opt. Lett.* **4** (1979) 21.
[26] J. Feinberg, R. W. Hellwarth, *Opt. Lett.* **5** (1980) 519.
[27] D. Z. Anderson, R. Saxena, M, Cronin-Golomb, *Phys. Rev. A*, **42** (1990) 3142.
[28] M. Cronin-Golomb, B. Fisher, J. O. White, A. Yariv, *IEEE J. Quantum Electorn.* **QE-20** (1984) 12.

[29] S. Weiss, S. Sternklar, B. Fisher, *Opt. Lett.* **12** (1987) 114.

[30]  P. Yeh, Introduction to Photorefractive Nonlinear Optics, John Wiley & Sons, New York, 1993. p. 315.